\documentclass[12pt]{article}

\usepackage[margin=1in]{geometry}
\usepackage{times}
\usepackage{hyperref}
\usepackage{url}
\usepackage{graphicx}
\usepackage{booktabs}
\usepackage{tabularx}
\usepackage{multirow}
\usepackage{array}
\usepackage{enumitem}
\usepackage{microtype}
\usepackage{xcolor}
\usepackage{setspace}
\usepackage[numbers]{natbib}

\hypersetup{
  colorlinks=true,
  linkcolor=blue,
  citecolor=blue,
  urlcolor=blue
}

\title{\textbf{Bastet: A Fine-Grained Expert-Labeled Dataset for DeFi Smart Contract Vulnerability Detection}}

\author{
  Wan-Hsuan Hsu\textsuperscript{1} \quad
  Wei-Hsin Wang\textsuperscript{1} \quad
  Cheng-Yu Liou\textsuperscript{1} \\[2pt]
  Ting-Rui Ke\textsuperscript{1} \quad
  Kentaroh Toyoda\textsuperscript{1} \\[4pt]
  \textsuperscript{1}AIFT \\[2pt]
  \small\texttt{\{alice.hsu, daky.wang, chengyu.liou\}@onesavie.com} \\[1pt]
  \small\texttt{\{kevin.ke, kentaroh.toyoda\}@aift.io}
}

\date{}

\begin{document}

\maketitle

\begin{abstract}
Smart contract vulnerabilities in Decentralized Finance (DeFi) protocols
resulted in over \$1.49 billion in confirmed losses in 2024 alone, across
192~incidents~\cite{immunefi2024}. As LLM-based vulnerability detection
emerges as a promising approach to address these threats, the quality of
evaluation datasets has become a critical bottleneck. Existing datasets suffer
from three fundamental problems: they are built on outdated Solidity versions
(e.g., v0.4) that no longer reflect modern DeFi contracts
\cite{xiao2025,feist2019,ghaleb2020}; they rely on automated or LLM-generated
annotations that introduce hallucination-driven label noise
\cite{ding2024,chen2025forge}; and they apply coarse single-layer labeling that
fails to capture the semantic complexity of real-world business logic
vulnerabilities \cite{feist2019,ghaleb2020,zheng2026dive,oliveira2024openscv}.
We present \textbf{Bastet}, an expert-labeled DeFi smart contract vulnerability
dataset that addresses all three problems through real-world audit findings
(2021--2024), human expert annotation with discussion-based consensus, and a
two-layer taxonomy of 46~Tags and 77~Subtags. Bastet comprises 4,402~findings
collected from 394~Code4rena competitive audit reports spanning April~2021 to
November~2024, of which 849~findings are fully annotated by white-hat security
researchers from the DeFiHackLabs community. All annotations are produced
through a two-annotator consensus workflow, ensuring label accuracy grounded
in real-world vulnerability root causes.
\end{abstract}

\section{Introduction}

The increasing complexity of DeFi protocols has outpaced the capabilities of
traditional static analysis tools. Chaliasos et al.\ showed that five mainstream
tools could only detect 11 out of 32~DeFi vulnerability types, with coverage
heavily concentrated on syntactic patterns such as reentrancy
\cite{chaliasos2024} --- leaving the majority of real-world business logic
vulnerabilities undetected. The consequences are severe: according to Immunefi,
the cryptocurrency industry suffered \$1.49~billion in losses from hacks and
exploits in 2024 alone~\cite{immunefi2024}. This gap has driven growing interest
in LLM-based vulnerability detection, which offers the semantic reasoning needed
to identify business logic flaws that static tools cannot reach
\cite{hossain2025,wei2025}.

However, the progress of LLM-based detection is constrained by a fundamental
problem: the quality of evaluation datasets. Without reliable, fine-grained
ground truth, it is impossible to fairly evaluate or improve detection methods.
Existing datasets fall short in three ways.

\textbf{First}, most datasets are built on outdated Solidity versions and are
heavily skewed toward a narrow set of syntactic vulnerabilities. Representative
datasets such as SmartBugs~\cite{feist2019} and SolidiFI~\cite{ghaleb2020} were
constructed primarily using contracts written in Solidity~v0.4, which are
significantly shorter and simpler than modern DeFi contracts. Reentrancy-focused
datasets such as ReentrancyStudy~\cite{zheng2023} and
SCRUBD~\cite{durieux2024scrubd} cover only one or two vulnerability types,
failing to reflect the diversity of vulnerabilities found in modern DeFi
protocols. Xiao et al.\ demonstrated that LLM evaluations relying on these
v0.4-based datasets fail to reflect real-world detection difficulty: when
evaluated on Solidity~v0.8 contracts, recall rates for certain vulnerability
types dropped to as low as 13\%~\cite{xiao2025}.

\textbf{Second}, datasets that rely on automated or LLM-generated annotations
carry hallucination-driven label noise. Ding et al.\ showed that existing
benchmarks suffer from severe label noise, causing models like GPT-4 to perform
close to random guessing under stricter evaluation settings~\cite{ding2024}.
More broadly, LLMs are known to produce plausible but factually incorrect
outputs when reasoning about subtle code semantics, a risk that is especially
pronounced for business logic vulnerabilities where root cause identification
requires deep domain expertise~\cite{chen2025forge}. A dataset labeled by an
LLM inherits these errors, making it an unreliable benchmark for evaluating
other LLM-based detectors.

\textbf{Third}, existing datasets apply coarse single-layer labeling. Labels
such as ``Reentrancy'' or ``Slippage'' do not distinguish between structurally
different root causes that require different detection strategies
\cite{feist2019,ghaleb2020,zheng2026dive,oliveira2024openscv}. For example, a
slippage vulnerability caused by a hardcoded parameter and one caused by a
missing protection entirely share the same top-level label but demand different
reasoning to detect.

To address these three challenges, we present \textbf{Bastet} --- an
expert-labeled DeFi smart contract vulnerability dataset built from real-world
competitive audit findings. Bastet makes the following contributions:

\begin{enumerate}[leftmargin=*, label=\arabic*.]
  \item \textbf{Expert-labeled Dataset}: 4,402~findings from 394~Code4rena
        audit reports (April~2021--November~2024), with 849~findings fully
        annotated by white-hat security researchers from the DeFiHackLabs
        community, accompanied by expert-written root cause summaries.

  \item \textbf{Two-Layer Taxonomy}: A structured classification scheme of
        46~Tags and 77~Subtags that distinguishes vulnerability root cause
        mechanisms from specific implementation flaws, enabling finer-grained
        and deep evaluation than any existing single-layer dataset.

  \item \textbf{Rigorous Annotation Process}: A two-annotator workflow where
        each finding is independently labeled by two white-hat researchers,
        with disagreements resolved through discussion-based consensus,
        ensuring that every label reflects genuine expert agreement on root
        cause.
\end{enumerate}

The remainder of this paper is organized as follows: Section~\ref{sec:related}
reviews limitations of existing datasets and annotation approaches;
Section~\ref{sec:dataset} describes the Bastet dataset construction;
Section~\ref{sec:analysis} presents dataset statistics and comparison with
existing datasets; Section~\ref{sec:availability} describes dataset
availability; Section~\ref{sec:conclusion} concludes the paper.

\section{Related Work}
\label{sec:related}

To motivate the design of Bastet, this section reviews the limitations of
existing smart contract vulnerability datasets and the risks of automated
annotation pipelines. Section~\ref{sec:related_datasets} surveys representative
datasets and identifies three recurring shortcomings;
Section~\ref{sec:related_hallucination} examines the specific problem of
hallucination-driven label noise in LLM-generated annotations.

\subsection{Limitations of Existing Smart Contract Vulnerability Datasets}
\label{sec:related_datasets}

The quality of vulnerability datasets directly determines how reliably detection
methods can be evaluated. Existing datasets fall into three categories, each
with significant limitations.

The first category consists of \textbf{automated bug injection datasets}.
SolidiFI~\cite{ghaleb2020} injects 9,369~synthetic bugs across seven
vulnerability types into 50~contracts written in Solidity~v0.4. While the
injection process is reproducible, the resulting vulnerabilities are
structurally simple and do not reflect the business logic complexity found in
real-world DeFi protocols. Evaluation results on SolidiFI therefore do not
generalize to production contracts.

The second category consists of \textbf{single-layer labeled datasets}.
SmartBugs~\cite{feist2019} provides 143~annotated Solidity contracts covering
syntax-level vulnerability types such as reentrancy, integer overflow, and
access control, also written primarily in Solidity~v0.4. Labels are applied at
the vulnerability-class level only, with no distinction between different root
cause patterns within the same class. Reentrancy-focused datasets such as
ReentrancyStudy~\cite{zheng2023} and SCRUBD~\cite{durieux2024scrubd} push scale
further but remain limited to one or two vulnerability types, failing to reflect
the diversity of vulnerabilities found in modern DeFi protocols. As Xiao et al.\
demonstrated, evaluating LLMs on these v0.4-based datasets significantly
underestimates real-world detection difficulty: recall rates for certain
vulnerability types dropped to as low as 13\% when evaluation was conducted on
modern Solidity~v0.8 contracts~\cite{xiao2025}.

The third category consists of \textbf{LLM-sourced datasets}. Recent work such
as LLM-SmartAudit~\cite{wei2025} sources findings from Code4rena but applies
binary labeling (vulnerable / not vulnerable) without a systematic taxonomy.
While the scale is large, binary labels cannot distinguish between structurally
different root causes, limiting the dataset's utility for training or evaluating
detection systems that need to reason about specific vulnerability patterns. More
broadly, DIVE~\cite{zheng2026dive} and OpenSCV~\cite{oliveira2024openscv} both
identify coarse single-layer labeling as a fundamental limitation of the field,
and call for finer-grained classification schemes that better reflect the
semantic diversity of real-world vulnerabilities.

\subsection{Hallucination Risk in LLM-generated Annotations}
\label{sec:related_hallucination}

As dataset construction increasingly turns to automated and LLM-based annotation
pipelines to achieve scale, a new concern has emerged: the reliability of
LLM-generated labels. While automated pipelines offer advantages in speed and
cost, they introduce hallucination-driven label noise that undermines dataset
quality. Ding et al.\ showed that existing benchmarks built with automated
annotation suffer from severe label noise, causing models like GPT-4 to perform
close to random guessing under stricter evaluation
settings~\cite{ding2024}. This result suggests that high reported accuracy
numbers on such benchmarks may not reflect genuine detection capability.

The problem is especially acute for smart contract vulnerability annotation.
Identifying the root cause of a DeFi business logic vulnerability requires deep
familiarity with protocol design, economic mechanisms, and common attack
patterns. LLMs are trained to replicate data distributions rather than validate
factual correctness, and tend to produce confident outputs even under
uncertainty~\cite{kalai2025hallucinate} --- making them prone to plausible but
incorrect root cause judgments precisely where domain expertise matters
most~\cite{chen2025forge}. A dataset annotated by an LLM therefore inherits
these errors and becomes an unreliable benchmark for evaluating other LLM-based
detectors. In contrast, human expert annotators with hands-on audit experience
can reason about vulnerabilities at the level of protocol semantics, and a
two-annotator consensus workflow provides a natural cross-check that automated
pipelines cannot replicate.

\section{The Bastet Dataset}
\label{sec:dataset}

Having established the limitations of existing datasets in
Section~\ref{sec:related}, this section describes how Bastet is constructed to
address them. Section~\ref{sec:datasource} explains the choice of data source;
Section~\ref{sec:taxonomy} presents the two-layer taxonomy design;
Section~\ref{sec:annotation} describes the expert annotation workflow.

\subsection{Data Source}
\label{sec:datasource}

The Bastet dataset is constructed from publicly available audit reports on
Code4rena~\cite{code4rena2021}, a competitive smart contract audit platform
where security researchers worldwide submit vulnerability findings in exchange
for financial rewards. Code4rena reports are particularly suitable as a data
source for two reasons. First, each finding has already been reviewed and
accepted by experienced contest judges before our annotation process begins,
providing a natural quality filter that eliminates low-quality or duplicate
submissions. Second, Code4rena findings cover a wide range of modern DeFi
protocols written in contemporary Solidity versions, including complex business
logic vulnerabilities that are absent from synthetic or legacy datasets.

We collected all publicly available audit reports on Code4rena spanning
April~2021 to November~2024, yielding 394~reports and a total of 4,402~findings
across diverse DeFi protocol categories including lending, derivatives, yield
aggregators, and decentralized exchanges.

\subsection{Taxonomy Design}
\label{sec:taxonomy}

The core contribution of the Bastet dataset is a two-layer classification
taxonomy developed iteratively with the DeFiHackLabs white-hat
community~\cite{defihacklabs2022}. Tags and Subtags were not defined in advance;
instead, they emerged from the annotation process itself. When annotators
observed two or more findings that shared a recurring vulnerability pattern not
yet captured by the existing taxonomy, they proposed a new Tag or Subtag,
discussed it with the team, and added it with a formal definition and associated
knowledge references. This bottom-up, iterative approach ensures that every
category in the taxonomy is grounded in at least two real-world observations,
rather than derived from theoretical assumptions.

\textbf{Tags (46~categories)} classify vulnerabilities by their root cause
mechanism. Each Tag is defined by a precise description, a set of knowledge
references, and a list of associated Subtags. For example, the \textit{Slippage}
Tag covers vulnerabilities where a contract fails to protect users from adverse
price movement during economic conversions.

\textbf{Subtags (77~categories)} identify the specific implementation flaw
within a given root cause. A single Tag may have multiple Subtags representing
structurally distinct patterns. For example, the \textit{Slippage} Tag contains
Subtags such as \texttt{minOut set to 0} (slippage protection is present but
disabled), \texttt{Missing minOut / maxAmount} (no protection exists at all),
and \texttt{Hardcoded Parameter} (protection exists but cannot adapt to market
conditions).

This two-layer design directly addresses the coarse single-layer labeling
problem identified in Section~\ref{sec:related}. A label such as ``Slippage''
at the Tag level alone cannot distinguish between these three patterns, which
require different detection logic and different remediation strategies. By
providing Subtag-level granularity alongside a one-sentence expert-written root
cause summary for each finding, Bastet enables more targeted prompt engineering,
supports fine-grained model evaluation, and allows detection tools to be
assessed at the level of specific vulnerability patterns rather than broad
categories.

\subsection{Annotation Process}
\label{sec:annotation}

\textbf{Annotation team.} All annotations are produced by white-hat security
researchers from the DeFiHackLabs community~\cite{defihacklabs2022}, a globally
recognized Web3 security community with over 4,000~members and more than
257~active white-hat researchers. Annotators have prior experience in Code4rena
audit competitions or on-chain incident response, providing the domain expertise
required to reason about DeFi business logic vulnerabilities.

\textbf{Workflow.} Each finding is independently labeled by two annotators. The
annotator reads the full finding report, assigns one or more Tags and
corresponding Subtags according to the taxonomy definitions, and writes a
one-sentence summary distilling the root cause of the vulnerability. If both
annotators assign the same labels, the annotation is accepted directly. If the
labels differ, the two annotators discuss the finding until consensus is reached.
No finding enters the dataset without full agreement on both Tag and Subtag
assignments.

This workflow ensures that every label in the dataset reflects genuine expert
consensus on root cause, rather than a single annotator's interpretation or an
automated label. The requirement for consensus rather than majority vote means
that ambiguous or borderline findings receive additional scrutiny before being
accepted, improving overall label reliability. By grounding all annotations in
human expert judgment, Bastet directly addresses the hallucination-driven label
noise identified in Section~\ref{sec:related_hallucination} --- a risk that
automated pipelines cannot eliminate.

\section{Dataset Analysis}
\label{sec:analysis}

With the dataset construction process described in Section~\ref{sec:dataset},
this section presents a quantitative analysis of the resulting Bastet dataset.
We report overall statistics, Tag and Subtag distributions, multi-label
characteristics, and a direct comparison with existing datasets.

\subsection{Dataset Overview}

The current release of the Bastet dataset contains 4,402~findings across
394~Code4rena audit reports. Of these, 849~findings have been fully annotated
with Tag, Subtag, and an expert-written root cause summary.
Table~\ref{tab:overview} summarizes the high-level dataset composition.

\begin{table}[h]
\centering
\caption{Bastet Dataset Overview}
\label{tab:overview}
\begin{tabular}{ll}
\toprule
\textbf{Attribute} & \textbf{Value} \\
\midrule
Total findings              & 4,402 \\
Annotated findings          & 849 \\
Audit reports               & 394 \\
Coverage period             & April 2021 -- November 2024 \\
Number of Tags              & 46 \\
Number of Subtags           & 77 \\
Findings with multiple Tags & 182 \\
Findings with multiple Subtags & 139 \\
Annotators                  & 3+ white-hat experts (2 per finding) \\
Peer-reviewed findings      & 100\% (all 849 annotated findings) \\
\bottomrule
\end{tabular}
\end{table}

\subsection{Tag Distribution}

Table~\ref{tab:tags} summarizes the distribution of Tags among the
849~fully annotated findings. This distribution reflects the currently annotated
subset only and should not be interpreted as representing the true prevalence of
vulnerability types across the full 4,402~findings.

\begin{table}[h]
\centering
\caption{Top-10 Tag Distribution (849 annotated findings)}
\label{tab:tags}
\begin{tabular}{clrr}
\toprule
\textbf{Rank} & \textbf{Tag} & \textbf{Count} & \textbf{Percentage} \\
\midrule
1  & DoS               & 176 & 20.7\% \\
2  & Input Validation  & 112 & 13.2\% \\
3  & Access Control    &  85 & 10.0\% \\
4  & Logic Error       &  72 &  8.5\% \\
5  & Accounting Error  &  71 &  8.4\% \\
6  & Arithmetic        &  60 &  7.1\% \\
7  & Slippage          &  54 &  6.4\% \\
8  & Reentrancy        &  40 &  4.7\% \\
9  & ERC20             &  38 &  4.5\% \\
10 & Oracle            &  37 &  4.4\% \\
\bottomrule
\end{tabular}
\end{table}

\subsection{Subtag Distribution}

Table~\ref{tab:subtags} summarizes the distribution of Subtags among the
849~fully annotated findings. As with the Tag distribution, this reflects the
current annotated subset only.

\begin{table}[h]
\centering
\caption{Top-10 Subtag Distribution (849 annotated findings)}
\label{tab:subtags}
\begin{tabular}{clrr}
\toprule
\textbf{Rank} & \textbf{Subtag} & \textbf{Count} & \textbf{Percentage} \\
\midrule
1  & Invalid Validation                      & 112 & 13.19\% \\
2  & Bad Condition                           &  79 &  9.30\% \\
3  & State Update Inconsistency              &  71 &  8.36\% \\
4  & Implementation Error                    &  51 &  6.01\% \\
5  & Incorrect Parameter                     &  50 &  5.89\% \\
6  & Centralization Risk                     &  39 &  4.59\% \\
7  & Price Manipulation / Arbitrage Opportunity & 33 & 3.89\% \\
8  & Violating CEI / Missing nonReentrant    &  32 &  3.77\% \\
9  & Front Run                               &  32 &  3.77\% \\
10 & Bypass Mechanism                        &  32 &  3.77\% \\
\bottomrule
\end{tabular}
\end{table}

The Subtag distribution demonstrates the fine-grained diversity of
implementation flaws captured in the dataset. Subtags such as State Update
Inconsistency, Price Manipulation, and Bypass Mechanism represent patterns that
require semantic reasoning to detect and are not covered by any existing
single-layer dataset.

\subsection{Multi-label Findings}

A distinctive characteristic of Bastet is its support for multi-label
annotation. Among the 849~annotated findings, 182~carry more than one Tag and
139~carry more than one Subtag. This reflects a real property of DeFi
vulnerabilities: a single finding often involves multiple interacting root
causes. For example, a vulnerability may combine an access control flaw that
allows unauthorized state changes with an accounting error in how those changes
are applied. Single-label datasets cannot represent this compound structure,
limiting their utility for training detection models that need to reason about
vulnerability interactions.

\subsection{Comparison with Existing Datasets}

Table~\ref{tab:comparison} provides a direct comparison between Bastet and
representative existing smart contract vulnerability datasets across key
dimensions relevant to LLM-based detection research.

\begin{table*}[t]
\centering
\caption{Comparison with Existing Smart Contract Vulnerability Datasets}
\label{tab:comparison}
\renewcommand{\arraystretch}{1.3}
\resizebox{\textwidth}{!}{%
\begin{tabular}{lllllll}
\toprule
\textbf{Dataset} & \textbf{Size} & \textbf{Solidity Version} & \textbf{Label Type} & \textbf{Annotation Source} & \textbf{Real-world} & \textbf{Root Cause Summary} \\
\midrule
SmartBugs~\cite{feist2019}
  & 143 contracts & v0.4 & Single-layer & Automated & Partial & No \\
SolidiFI~\cite{ghaleb2020}
  & 50 contracts, 9,369 injected bugs & v0.4 & Single-layer & Automated injection & No & No \\
ReentrancyStudy~\cite{zheng2023}
  & 230,548 contracts & Mixed & Single-layer (reentrancy only) & Automated + manual verification & Yes & No \\
SCRUBD~\cite{durieux2024scrubd}
  & Not specified & Mixed & Single-layer (2 types) & Crowdsourced + expert & Yes & No \\
LLM-SmartAudit~\cite{wei2025}
  & 6,454 contracts & Mixed & Binary (vuln / not vuln) & Automated & Partial & No \\
\textbf{Bastet (ours)}
  & \textbf{849 annotated (4,402 total)} & \textbf{Mixed Solidity versions (2021--2024)} & \textbf{Two-layer (46 Tags + 77 Subtags)} & \textbf{3+ white-hat experts + consensus} & \textbf{Yes (Code4rena)} & \textbf{Yes} \\
\bottomrule
\end{tabular}%
}
\end{table*}

Bastet is the only dataset in this comparison that combines real-world audit
findings, modern Solidity version coverage, two-layer fine-grained labels,
expert community annotation with consensus review, and expert-written root cause
summaries. Each of these properties individually addresses a known limitation of
prior datasets; together they provide a qualitatively different resource for the
research community.

\section{Dataset Availability}
\label{sec:availability}

To support reproducibility and community use, this section describes how the
Bastet dataset can be accessed and what it contains.

The Bastet dataset is publicly available at
\href{https://www.kaggle.com/competitions/onesavie-bastet/overview}{here}. The dataset
is distributed as a zip archive containing the smart contract codebase, audit
report findings, and a corresponding label CSV file for each repository. The
full labeled dataset will be released publicly upon the conclusion of the
competition. The dataset is released under the Creative Commons
Attribution-NonCommercial 4.0 International (CC~BY-NC~4.0) license.

\section{Conclusion}
\label{sec:conclusion}

We presented Bastet, an expert-labeled DeFi smart contract vulnerability dataset
that directly addresses three fundamental limitations of existing datasets. To
tackle outdated Solidity coverage, Bastet is built from 394~real-world
Code4rena audit reports spanning April~2021 to November~2024. To eliminate
hallucination-prone automated annotations, all 849~annotated findings are
labeled by white-hat security researchers from the DeFiHackLabs community
through a two-annotator consensus workflow. To overcome coarse single-layer
labeling, Bastet introduces a two-layer taxonomy of 46~Tags and 77~Subtags that
captures both root cause mechanisms and specific implementation flaws. We hope
Bastet serves as a reliable foundation for training and evaluating LLM-based
DeFi vulnerability detection systems. Future work includes expanding annotation
coverage to the full 4,402~findings and extending the taxonomy as new
vulnerability patterns emerge.

\bibliographystyle{unsrtnat}
\bibliography{references}

@misc{immunefi2024,
  author       = {{Immunefi}},
  title        = {Crypto Losses in 2024},
  year         = {2024},
  howpublished = {Immunefi Research Report},
  url          = {https://immunefi.com/research/}
}

@inproceedings{chaliasos2024,
  author    = {Chaliasos, Stefanos and Charalambous, Marcos A. and Zhou, Liyi and Galanopoulou, Raghavendra and Gervais, Arthur and Mitropoulos, Dionysis and Livshits, Benjamin},
  title     = {Smart Contract and {DeFi} Security Tools: Do They Meet the Needs of Practitioners?},
  booktitle = {Proceedings of the 46th IEEE/ACM International Conference on Software Engineering (ICSE)},
  year      = {2024},
  doi       = {10.1145/3597503.3623302}
}

@inproceedings{hossain2025,
  author    = {Hossain, S. M. Mehedi and Altarawneh, Ammar and Roberts, Jerry},
  title     = {Leveraging Large Language Models and Machine Learning for Smart Contract Vulnerability Detection},
  booktitle = {Proceedings of the 2025 IEEE 15th Annual Computing and Communication Workshop and Conference (CCWC)},
  year      = {2025},
  url       = {https://arxiv.org/abs/2501.02229}
}

@article{wei2025,
  author  = {Wei, Zhiyuan and Sun, Jing and Sun, Yuhang and Liu, Yi and Wu, Defeng and Zhang, Zheng and Zhang, Xiaofei and Li, Meng and Liu, Yang and Li, Chunrong and Wan, Ming and Dong, Jin Song and Zhu, Liming},
  title   = {Advanced Smart Contract Vulnerability Detection via {LLM}-Powered Multi-Agent Systems},
  journal = {IEEE Transactions on Software Engineering},
  volume  = {51},
  number  = {10},
  pages   = {2830--2846},
  year    = {2025},
  doi     = {10.1109/TSE.2025.3597319}
}

@misc{xiao2025,
  author        = {Xiao, Zexin and Wang, Qingyuan and Pearce, Hammond and Chen, Sze Yiu},
  title         = {Logic Meets Magic: {LLMs} Cracking Smart Contract Vulnerabilities},
  year          = {2025},
  eprint        = {2501.07058},
  archivePrefix = {arXiv},
  url           = {https://arxiv.org/abs/2501.07058}
}

@inproceedings{feist2019,
  author    = {Ferreira, Jo{\~a}o F. and Cruz, Pedro and Durieux, Thomas and Abreu, Rui},
  title     = {{SmartBugs}: A Framework to Analyze Solidity Smart Contracts},
  booktitle = {Proceedings of the 35th IEEE/ACM International Conference on Automated Software Engineering (ASE)},
  year      = {2020}
}

@inproceedings{ghaleb2020,
  author    = {Ghaleb, Asem and Pattabiraman, Karthik},
  title     = {How Effective Are Smart Contract Analysis Tools? {Evaluating} Smart Contract Static Analysis Tools Using Bug Injection},
  booktitle = {Proceedings of the 29th ACM SIGSOFT International Symposium on Software Testing and Analysis (ISSTA)},
  year      = {2020},
  doi       = {10.1145/3395363.3397385}
}

@inproceedings{zheng2023,
  author    = {Zheng, Zibin and Chen, Neng and Ye, Jianzhong and Chen, Jiachi and Hu, Zhijie and Wu, Lianghao and Luo, Weizhe and Jiang, Wenjia and Xie, Zhiying and Wang, Yanlin},
  title     = {Turn the Rudder: A Beacon of Reentrancy Detection for Smart Contracts on {Ethereum}},
  booktitle = {Proceedings of the 45th IEEE/ACM International Conference on Software Engineering (ICSE)},
  year      = {2023}
}

@misc{ding2024,
  author        = {Ding, Yuqing and Fu, Yifan and Ibrahim, Omar and Sitawarin, Chawin and Chen, Xinyun and Alomair, Basel and Wagner, David and Ray, Baishakhi and Chen, Yizheng},
  title         = {Vulnerability Detection with Code Language Models: How Far Are We?},
  year          = {2024},
  eprint        = {2403.18624},
  archivePrefix = {arXiv},
  url           = {https://arxiv.org/abs/2403.18624}
}

@misc{chen2025forge,
  author        = {Chen, Tao and others},
  title         = {{FORGE}: An {LLM}-Driven Framework for Large-Scale Smart Contract Vulnerability Dataset Construction},
  year          = {2025},
  eprint        = {2506.18795},
  archivePrefix = {arXiv},
  url           = {https://arxiv.org/abs/2506.18795}
}

@article{zheng2026dive,
  author  = {Zheng, Huashan and others},
  title   = {{DIVE}: A Multi-Label Smart Contract Vulnerability Dataset},
  journal = {Scientific Data},
  year    = {2026},
  url     = {https://www.nature.com/articles/s41597-026-07025-5}
}

@article{oliveira2024openscv,
  author  = {Oliveira, Jo{\~a}o and others},
  title   = {{OpenSCV}: An Open Hierarchical Taxonomy for Smart Contract Vulnerabilities},
  journal = {Empirical Software Engineering},
  year    = {2024},
  doi     = {10.1007/s10664-024-10446-8}
}

@misc{durieux2024scrubd,
  author        = {Durieux, Thomas and others},
  title         = {{SCRUBD}: Smart Contracts Reentrancy and Unhandled Exceptions Vulnerability Dataset},
  year          = {2024},
  eprint        = {2412.09935},
  archivePrefix = {arXiv},
  url           = {https://arxiv.org/abs/2412.09935}
}

@misc{code4rena2021,
  author       = {{Code4rena}},
  title        = {{Code4rena}: Competitive Smart Contract Audit Platform},
  year         = {2021},
  howpublished = {Online},
  url          = {https://code4rena.com/}
}

@misc{defihacklabs2022,
  author       = {{SunWeb3Sec}},
  title        = {{DeFiHackLabs}: {Web3} Security Incident Database},
  year         = {2022},
  howpublished = {GitHub Repository},
  url          = {https://github.com/SunWeb3Sec/DeFiHackLabs}
}

@misc{kalai2025hallucinate,
  author        = {Kalai, Adam Tauman and Nachum, Ofir},
  title         = {Why Language Models Hallucinate},
  year          = {2025},
  eprint        = {2509.04664},
  archivePrefix = {arXiv},
  url           = {https://arxiv.org/abs/2509.04664}
}

\end{document}